\documentclass[twocolumn, twocolappendix]{aastex631}

\usepackage{comment}
\usepackage{color}
\usepackage{appendix}
\usepackage{amsmath}
\usepackage{ulem}
\newcommand\m{\mathrm}

\newcommand {\kms}{\mbox{km~s$^{-1}$}}

\newcommand\ax{{AX J1745.6$-$2901}}
\newcommand\maxi{{MAXI J1744$-$294}}
\newcommand\chandra{{\it Chandra}}
\newcommand\xmm{{\it XMM-Newton}}

\newcommand\nustar{{\it NuSTAR}}

\newcommand\xrism{{\it XRISM}}

\usepackage{xcolor}
\usepackage{ulem}

\begin{document}


\title{Distinct Velocity Components in the Absorption Lines of the Neutron Star X-ray Binary AX J1745.6–2901
}

\shorttitle{Distinct Velocity Components in the Absorption Lines of AX J1745.6–2901}
\shortauthors{Matsunaga et al.}

\correspondingauthor{Kai Matsunaga}
\email{matsunaga.kai.i47@kyoto-u.jp}
\correspondingauthor{Maxime Parra}
\email{maxime.parrastro@gmail.com}

\author[0009-0003-0653-2913]{Kai Matsunaga}
\affiliation{Department of Physics, Graduate School of Science, Kyoto University, Kitashirakawa Oiwake-cho, Sakyo-ku, Kyoto 606-8502, Japan}

\author[0009-0003-8610-853X]{Maxime Parra}
\affiliation{Department of Physics, Ehime University, 2-5, Bunkyocho, Matsuyama, Ehime 790-8577, Japan}

\author[0009-0003-9261-2740]{Yutaro Nagai}
\affiliation{Department of Physics, Graduate School of Science, Kyoto University, Kitashirakawa Oiwake-cho, Sakyo-ku, Kyoto 606-8502, Japan}

\author[0000-0003-1244-3100]{Teruaki Enoto} 
\affiliation{Department of Physics, Graduate School of Science, Kyoto University, Kitashirakawa Oiwake-cho, Sakyo-ku, Kyoto 606-8502, Japan}

\author[0000-0002-9099-5755]{Yoshitomo Maeda}
\affiliation{Institute of Space and Astronautical Science (ISAS), Japan Aerospace Exploration Agency (JAXA), 3-1-1 Yoshinodai, Chuo-ku, Sagamihara,
Kanagawa 252-5210, Japan}

\author[0000-0001-6665-2499]{Takayuki Hayashi}
\affiliation{Department of Physics, Graduate School of Science, Kyoto University, Kitashirakawa Oiwake-cho, Sakyo-ku, Kyoto 606-8502, Japan}

\author[0000-0002-6126-7409]{Shifra Mandel}
\affiliation{Columbia Astrophysics Laboratory, Columbia University, New York, NY 10027, USA}

\author[0000-0002-9709-5389]{Kaya Mori} 
\affiliation{Columbia Astrophysics Laboratory, Columbia University, New York, NY 10027, USA}

\author[0000-0003-4580-4021]{Hideki Uchiyama}
\affiliation{Faculty of Education, Shizuoka University, 836 Ohya, Suruga-ku, Shizuoka, Shizuoka 422-8529, Japan}

\author[0000-0003-1130-5363]{Masayoshi Nobukawa}
\affiliation{Faculty of Education, Nara University of Education, Nara, 630-8502, Japan}

\author[0000-0002-5092-6085]{Hiroya Yamaguchi}
\affiliation{Institute of Space and Astronautical Science (ISAS), Japan Aerospace Exploration Agency (JAXA), 3-1-1 Yoshinodai, Chuo-ku, Sagamihara,
Kanagawa 252-5210, Japan}

\author[0000-0001-8195-6546]{Megumi Shidatsu} 
\affiliation{Department of Physics, Ehime University, 2-5, Bunkyocho, Matsuyama, Ehime 790-8577, Japan}

\author[0000-0002-6797-2539]{Ryota Tomaru}
\affiliation{Faculty of Natural Sciences,
National Institute of Technology (KOSEN), Kure College, 2-2-11 Aga-minami, Kure, Hiroshima,  737-8506, Japan}

\begin{abstract}
\textcolor{black}{Accretion disks in X-ray binaries regulate mass transfer onto compact objects and drive radiative and kinetic feedback to their surroundings.} Here we report X-ray spectroscopy of the eclipsing neutron star low-mass X-ray binary \ax{} with \xrism{}/Resolve.
The phase-averaged Fe~\textsc{xxvi} Ly$\alpha$ absorption profile exhibits two absorption minima with relative depths that are inconsistent with the theoretical Ly$\alpha_1$/Ly$\alpha_2$ doublet ratio expected from a single velocity component.
We demonstrate that this profile is well described by two discrete velocity components: a blueshifted component at $v \simeq -160$~km~s$^{-1}$ and a redshifted component at $v \simeq +590$~km~s$^{-1}$. The significance of the redshifted component is more than $3\sigma$ based on a Monte-Carlo calculation. This velocity structure persists across orbital phases, disfavoring a localized origin such as a bulge or dip.
The blueshifted component, well below the outer-disk escape velocity, is consistent with a slow outflow or disk atmosphere. 
\textcolor{black}{The redshifted absorber can be explained either by infalling gas from a failed wind or by a gravitational redshift, and the present data cannot rule out either possibility. Regardless of its origin, the redshifted component is kinematically separate from the disk atmosphere and outflow. The absence of absorption at intermediate velocities further indicates a genuinely bimodal velocity distribution rather than the two ends of a single continuous flow, offering a new view of the absorbing-gas kinematics.}

\end{abstract}


\section{Introduction} \label{sec:intro}

\textcolor{black}{Accretion disks are ubiquitous in the Universe and play a central role in mass transfer, energy release, and feedback to their surroundings.} In low-mass X-ray binaries (LMXBs), a compact object accretes from a low-mass companion through Roche-lobe overflow, forming an accretion disk from which outflows are commonly launched. When such a system is viewed at high inclination, the line of sight passes close to the disk plane and intercepts highly ionized absorbing gas: a disk atmosphere, a nearly static layer above the disk surface with small line-of-sight velocity, and, in many cases, a disk wind seen as blueshifted absorption \citep{ponti_2012a, diaztrigo_2016}.

\textcolor{black}{Recent high-resolution X-ray spectroscopy has revealed complex velocity structures in several LMXBs. GX~13+1 shows a stratified wind with multiple blueshifted components \citep{audard_2025}, while 4U~1630$-$472 and MAXI~J1305$-$704 exhibit both blueshifted and redshifted absorption, interpreted as possible failed winds \citep{miller_2025,miller_2014}. A redshifted component in 4U~1916$-$053 has instead been discussed as a gravitationally redshifted disk atmosphere, because it remains at a nearly constant velocity despite long-term luminosity changes \citep{trueba_2020}. Redshifted absorption has also been reported in the eclipsing LMXB XTE~J1710$-$281, where a significant redshift of the highly ionized absorption lines was interpreted as possible gravitational redshift rather than inflow, because no clear inverse P-Cygni profile was detected \citep{trueba_2022}. These results suggest that disk atmospheres, winds, and infalling material may coexist in high-inclination LMXBs, but it remains unclear whether they are physically distinct structures or different stages of a continuous mass-circulation flow through the disk \citep{munoz-darias_2026a}}.

\textcolor{black}{To disentangle these complex velocity structures, absorption features from highly ionized iron species, particularly Fe \textsc{xxv} and Fe \textsc{xxvi}, provide the most direct constraints. These lines are commonly observed owing to the high cosmic abundance of iron and their presence in the 6--7 keV band, where attenuation by continuum absorption is modest. However, with previous X-ray instruments, including \chandra{} High Energy Transmission Grating (HETG), the fine structures of these line complexes were often blended. 
Here, observations with \xrism{}/Resolve overcome this limitation.}
\xrism{}/Resolve achieves an energy resolution of $\sim$5~eV at 6~keV, enabling the spectral separation of the Fe~\textsc{xxvi} Ly$\alpha_1$ and Ly$\alpha_2$ doublet \citep[e.g.,][]{audard_2025,trigo_2026}.

\textcolor{black}{AX~J1745.6$-$2901 is an eclipsing LMXB located in the vicinity of Sgr~A* at the Galactic center, with an inclination angle of approximately $70^\circ$ and an orbital period of $P_{\rm orb}\simeq3.0\times10^4$~s \citep[e.g.,][]{maeda_1996, ponti_2017, jin_2018, tanaka_2026}.}
Previous observations confirmed that the compact object in \ax\ is a neutron star, based on the detection of 
type-I X-ray bursts \citep{maeda_1996, degenaar_2009}. 
\textcolor{black}{In the soft state, the spectrum of \ax{} is dominated by thermal emission with at most a weak Comptonization (corona) component \citep{ponti_2018}, while IXPE has revealed the highest polarization degree yet measured for a NS-LMXB, rising sharply during eclipses and dips \citep{mikusincova_2025}. These results indicate that a substantial fraction of the soft-state emission is scattered light arising in the disk wind/atmosphere in this high-inclination geometry.}
\citet{trueba_2022} detected Fe~\textsc{xxv} and Fe~\textsc{xxvi} absorption lines with \chandra{}/HETG and reported a line width of $\sigma=100^{+40}_{-30}$~km~s$^{-1}$ and a redshift of $v=+270^{+240}_{-230}$~km~s$^{-1}$.  
\citet{tanaka_2026} reported the spectrum obtained with \xrism{} in the performance verification phase in 2024. They measured the widths and the shifts of absorption lines of Fe \textsc{xxvi} and Ni \textsc{xxviii} as $\sigma=110^{+40}_{-30}$~km~s$^{-1}$ and $v_{\rm out}=-160^{+50}_{-70}$~km~s$^{-1}$ using a single ionized absorber; however, because \ax\ lay outside the nominal field of view of Resolve in that observation, the photon statistics and signal-to-background ratio were very limited, preventing a detailed investigation of the fine structures.

In this paper, we report the spectral analysis of a new \xrism{}/Resolve observation of AX~J1745.6--2901 obtained in 2025. 
\textcolor{black}{The observation lasted 128~ks, with a net exposure of 71~ks, and covered about four orbital cycles. This enabled us to investigate both the fine velocity structure and its orbital-phase dependence, because \ax{} was brighter and located within the Resolve field of view, unlike in the 2024 observation.}


\textcolor{black}{We note that the same XRISM observation was previously analyzed using the full Resolve array, and the Fe-band features were attributed mainly to MAXI~J1744$-$294 \citep{chatterjee_2025}. Subsequent source-disentangling analysis showed that MAXI~J1744$-$294 exhibits emission features, while the strong absorption features arise from the \ax{}-dominated region \citep{parra_2026, parra_2026a}. }

\section{Observation and Data Reduction} \label{sec:obs}

The \xrism{} observation 
analyzed in this study was conducted between 2025-03-03 UT 11:30 and 2025-03-05 UT 00:21 with a net exposure of 71 ks as Director's Discretionary Time, during a multi-wavelength campaign investigating the evolution of the Galactic center black hole transient MAXI J1744-294/Swift J174540.2-290037 \citep[][ObsID: 901002010]{mandel_2025, parra_2026, parra_2026a}. While the primary target was the outburst of \maxi{}, \ax{} was located on the edge of the FoV of Resolve, yielding higher photon statistics than the 2024 observation. 

\begin{figure}
\centering
\includegraphics[width=1\linewidth]{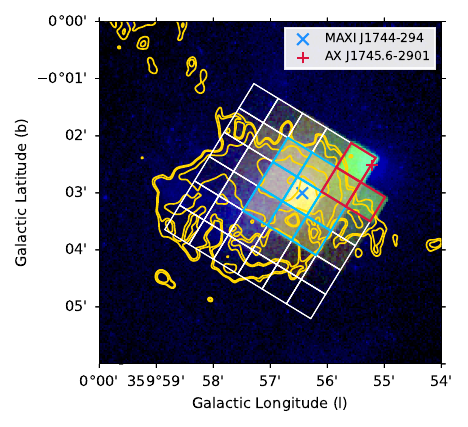}
\caption{X-ray RGB image obtained with \xrism{}/Resolve and \chandra{}/ACIS. The red and green channels correspond to the 2.0--5.0 keV and 5.0--10.0 keV bands observed with Resolve, respectively, while the blue channel represents the 2.0--7.0 keV band observed with ACIS\textcolor{black}{, in which the diffuse emission arises from mainly Sgr A East supernova remnant and GCXE}. \textcolor{black}{The gold contours show the VLA 4.8 GHz data from the NRAO archive (\url{https://www.vla.nrao.edu/astro/archive/pipeline/position/J174535.6-285839/}), indicating the position of Sgr A east supernova remnant with a radio shell of $3\farcm5 \times 2\farcm5$ \citep{ekers_1975}.
The positions of \ax{} 
and \maxi{} 
are marked. The red and cyan regions indicate the \ax{} and \maxi{} extraction regions for the spectral analysis.
}}
\label{fig:map}
\end{figure}

As shown in Figure~\ref{fig:map}, the Resolve FoV of 
$3\farcm1\times3\farcm1$ with a $6\times6$ pixel array is contaminated by several X-ray sources other than \ax{}. 
The black hole LMXB \maxi{}, 
the primary target of this observation, is separated from \ax\ by $\sim1\farcm3$
.  The supernova remnant Sgr~A~East 
 whose X-ray spectrum is dominated by over-ionized plasma \citep{xrismcollaboration_2025}, is separated from \ax\ by $\sim1\farcm5$. 
Additional contamination is presented by the diffuse Galactic center X-ray emission (GCXE), 
whose origin is still not well understood \citep{koyama_2018}. To isolate the emission from \ax, we defined 
source and background regions, indicated by red and cyan in Figure~\ref{fig:map}, respectively, and modeled the two spectra simultaneously. We refer to \citet{parra_2026} for details of the methodology.

\textcolor{black}{For each extraction region and source component, we generated spectra, RMFs, and ARFs using HEASOFT v6.36 and the Resolve Calibration Database (CalDB) v12. We used only High-primary events. The ARFs were calculated with \texttt{xaxmaarfgen}, adopting point-source responses for \ax{} and \maxi{}, a Chandra 6.6--6.8~keV band image as a spatial template for Sgr~A East, and a spatially flat model for the GCXE. We corrected the known HEASOFT v6.36/CALDB v12 region-handling issue in \texttt{xaxmaarfgen} by following the XRISM Science Data Center recommendation (details are given in Appendix~\ref{app:arf}). Even after this response modeling, small residual cross-contamination between the \ax{} and \maxi{} extraction regions remained. We therefore introduced multiplicative constants to correct for the mutual cross-contamination between the \ax{} and \maxi{} extraction regions. The diffuse contributions from Sgr~A East and the GCXE were weak enough that analogous corrections were unnecessary. The best-fit cross-contamination factors are both $\sim0.8$ (Table~\ref{tab:bestfit}).}

\begin{figure*}[t]
\centering
\includegraphics[width=\linewidth]{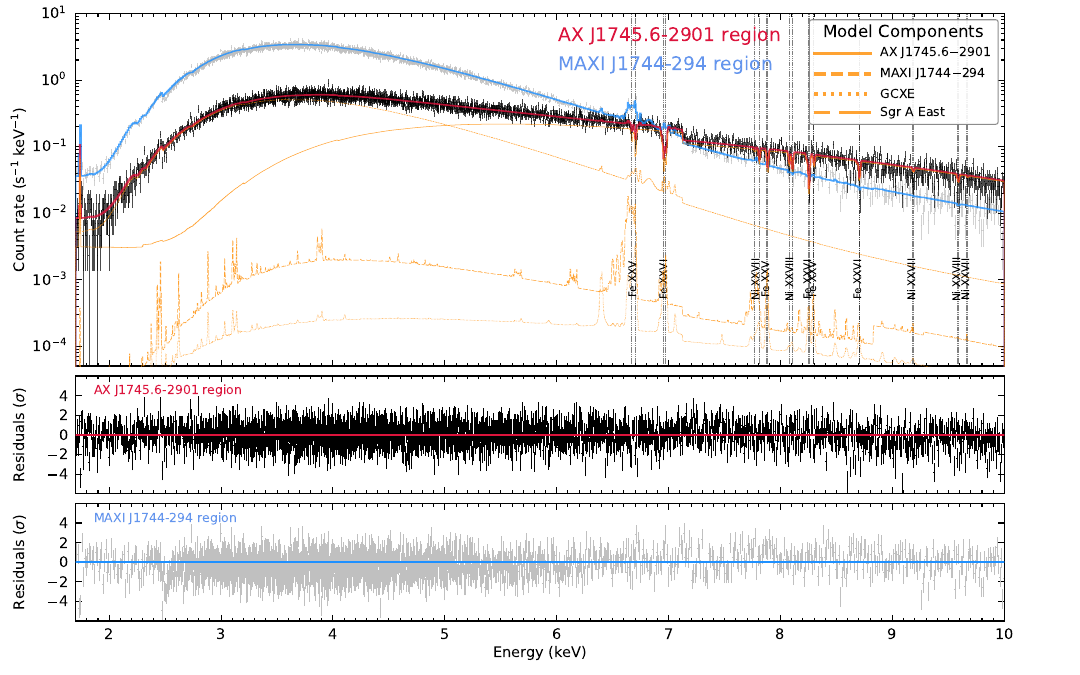}
\caption{\textcolor{black}{Phase-averaged XRISM/Resolve spectra of the \ax{} extraction region and the \maxi{} extraction region, fitted simultaneously to account for contamination in the Resolve field of view. The \ax{} data and best-fit total model are shown in black and red, respectively, while the \maxi{} data and best-fit total model are shown in silver and blue, respectively. The best-fit model includes two velocity absorption components in \ax{}. Orange curves show the model contributions to the \ax{} region spectrum from \ax{} (solid), \maxi{} (dashed), Sgr~A East (long-dashed), and the Galactic center X-ray emission (GCXE; dotted). The labeled vertical markers indicate the main absorption features included in the \ax{} model. The lower panels show the residuals in units of (data$-$model)/error.
}}
\label{fig:spec}
\end{figure*}

\begin{figure}
    \centering
    \includegraphics[width=\linewidth]{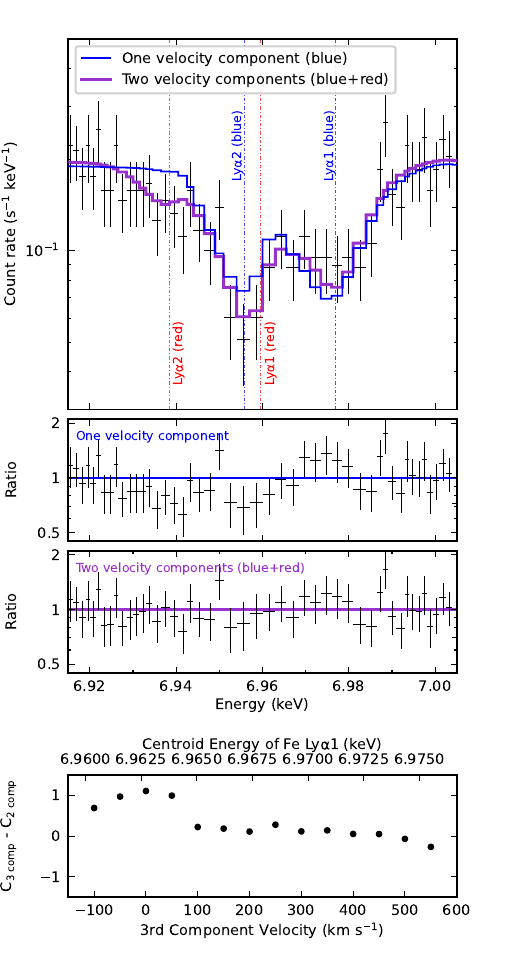}
    \caption{Enlarged view of the Fe XXVI Ly$\alpha$ absorption region in the phase-averaged \xrism{}/Resolve spectrum of AX J1745.6–2901. Top panel: The spectrum fitted with a single-velocity absorption model (thin purple) and a two-velocity-component model (thick purple). The vertical dashed lines indicate the rest-frame energies of the Fe \textsc{xxvi}~Ly$\alpha_1$ and $\alpha_2$ transitions, shifted according to the best-fit Doppler velocities of the two components. Middle and lower panels: Ratios of the data to the best-fit models for the single- and two-velocity-component cases, respectively. The two-component model significantly reduces the residuals around the Fe \textsc{xxvi} Ly$\alpha$ doublet. Bottom panel: $\Delta$C-statistic improvement obtained by introducing an additional third velocity component as a function of its Doppler velocity. For each trial velocity, only the column density was allowed to vary while the velocity width was fixed to that of the other components. No statistically significant improvement is found at intermediate velocities, supporting the presence of two discrete velocity components rather than a continuous velocity distribution.}
    \label{fig:spec_zoom}
\end{figure}

\begin{figure*}
    
\includegraphics[width=0.65\linewidth]{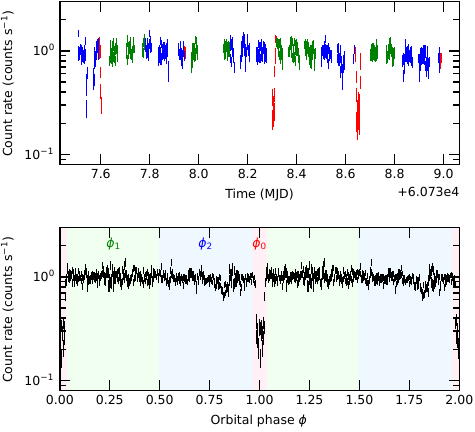}
\includegraphics[width=0.335\linewidth]{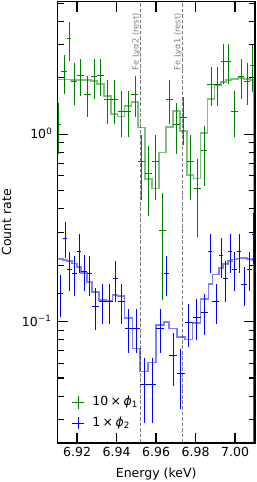}
    \caption{\textcolor{black}{Left: The X-ray light curve (top) and folded light curve (bottom) of the \ax{} region. We assumed the orbital period of 30063~s, as reported from the 2024 observation \citep{tanaka_2026}. Based on the folded light curve, we excluded the eclipse interval ($\phi_0: \phi = 0.96$--$0.04$) and divided the remaining out-of-eclipse orbit into two natural phase bins: $\phi_1$ ($\phi = 0.04$--$0.50$) and $\phi_2$ ($\phi = 0.50$--$0.96$). The colors in the light curves correspond to these phase definitions \textcolor{black}{in the folded light curve}. Right: The scaled, zoomed spectra and the best-fit models in $\phi_1$ and $\phi_2$ around the Fe \textsc{xxvi} Ly$\alpha$. The $\phi_1$ spectrum is multiplied by a factor of 10 for clarity.}}
    \label{fig:lc}
\end{figure*}

\section{Spectral Analysis} \label{sec:ana}

\subsection{Phase-Averaged Spectral Analysis}


As mentioned in Section \ref{sec:obs}, contamination is expected in the \ax{} region from Sgr A East \citep{xrismcollaboration_2025}, the GCXE, and \maxi{}
. Therefore, to properly account for contamination, we simultaneously fitted the spectra of AX~J1745.6$-$2901 and the background components extracted from the two regions shown in Figure~\ref{fig:map}. The description of the other sources than \ax{} is detailed in Appendix~\ref{app:bgd}.

Here we focus on the modeling of the \ax\ component in the spectra. Figure \ref{fig:spec} shows the phase-averaged Resolve spectrum of the AX J1745.6-2901 region (black)\textcolor{black}{, including the eclipse phases, which occupy less than 10\% of the orbit}. For the continuum emission, following \citet{tanaka_2026}, we applied a blackbody plus disk-blackbody model, corresponding to the blackbody emissions from the neutron star surface and the accretion disk. In this observation, the high-level contamination from \maxi, particularly below $E\sim6$~keV, prevents the determination of their temperatures. Therefore, we fixed them to $kT_{bbody}=2.00$ and $kT_{disk}=1.14$ keV, respectively, obtained with \xrism{}/Xtend observation in 2024 \citep{tanaka_2026}, when \ax{} was not contaminated by \maxi{}. Note that the temperature of the continuum emissions hardly affect the absorption line profiles that we focused on in this study.

We modeled the highly ionized Fe and Ni absorption lines with the \texttt{voigtabs} model in XSPEC.
The ions and transitions included in the model are listed in Table~\ref{tab:atomic}.
The natural widths and oscillator strengths were fixed to laboratory values from the NIST Atomic Spectra Database \citep{nist_asd_2024}, allowing each ionic series of absorption lines to be described by only two free parameters: the column density $N_{\rm ion}$ and the turbulent velocity $\sigma_{\rm turb}$. See Appendix~\ref{app:voi} for details. The Doppler velocity $v_{\rm shift}$ and $\sigma_{\rm turb}$ in all \texttt{voigtabs} components were tied, assuming that all ions originate from the same kinematic component. We further convolved the model with a Gaussian smoothing component (\texttt{gsmooth}) to account for additional effects by bulk broadening velocity $\sigma_{\rm bulk}$ that are not explicitly included in the Voigt profiles, e.g., orbital motion. We also account for the interstellar absorption using \texttt{TBabs} model \citep{wilms_2000}. The initial spectral model was therefore
\begin{equation}
    \texttt{TBabs}*\texttt{gsmooth}*\texttt{vashift}*\Pi~ \texttt{voigtabs}*(\texttt{bbody+diskbb}).
    \label{eq:model}
\end{equation}

The model provides a good description of the broadband spectra of both the \ax{} region and \maxi{} region, including the Fe~\textsc{xxv}, Fe~\textsc{xxvi}, Ni~\textsc{xxvii}, and Ni~\textsc{xxviii} absorption features in the \ax{} component.
We did not detect a signature of Fe~\textsc{xxiv} component ($N_{\rm ion}<6\times10^{16}~{\rm cm ^{-2}}$ with 3$\sigma$ upper limit). The value of $N_{\rm H}$ was $N_{\rm H} = (24.3\pm1.0)\times10^{22}$~cm$^{-2}$. The normalizations of the blackbody and disk blackbody\footnote{Their definitions are described in the note of table 1.} in this observation were $\rm Norm_{\rm bbody}=6.2\pm0.1\times10^{-3}$ and $\rm Norm_{\rm disk}=24\pm1$, respectively. 
\textcolor{black}{The absorbed 3--6~keV flux was $1.00^{+0.02}_{-0.02}\times10^{-10}$~erg~cm$^{-2}$~s$^{-1}$, roughly consistent with and slightly higher than that reported by \citet{tanaka_2026}. The model extrapolated over the 0.01--100~keV band gives an unabsorbed bolometric luminosity of $L_{\rm bol}=1.40\times10^{37}{\rm~erg~s^{-1}}(d/8~{\rm kpc})^2$ (Table~\ref{tab:bestfit}).} 


However, the 6.9--7.0~keV band of the \ax\ region spectrum, dominated by Fe~\textsc{xxvi} Ly$\alpha$ transitions, shows significant residuals as shown in Figure~\ref{fig:spec_zoom}. Although the intrinsic intensity ratio of the Fe~\textsc{xxvi} Ly$\alpha$ doublet is expected to satisfy Ly$\alpha_1 ~(E\sim6.98~\rm keV)$ $>$ Ly$\alpha_2~(E\sim6.96~\rm keV)$ 
, the absorption feature at 6.96~keV appears deeper than the one at 6.98~keV. 

\textcolor{black}{
As a possible explanation for the anomalous Fe~\textsc{xxvi} Ly$\alpha_1$/Ly$\alpha_2$ depth ratio, we tested whether an additional Fe~\textsc{xxvi} Ly$\alpha$ emission component could account for the observed profile. This possibility is motivated by recent XRISM observations of the neutron-star dipper 4U~1624$-$49, where anomalous Fe~\textsc{xxvi} Ly$\alpha_1$/Ly$\alpha_2$ ratios, including inversions, were discussed in terms of optical-depth effects and re-emission in photoionized plasma \citep{trigo_2026}. Our test also empirically accounts for possible residual uncertainties in the contamination model, including emission from \maxi{} or diffuse thermal plasma components. We introduced Fe~\textsc{xxvi} Ly$\alpha_1$ and Ly$\alpha_2$ emission lines with a fixed 2:1 intensity ratio, and allowed their normalization, Doppler velocity, and broadening to vary, both with and without the \texttt{voigtabs} absorption applied. We scanned the Doppler velocity from $v=-1000$ to $+1000~\kms$ in steps of $\Delta v=50~\kms$, corresponding to $\Delta E\simeq1.2$~eV at 7~keV, smaller than the Resolve energy resolution. The most favorable solution was found at $v\simeq+20~\kms$ with $\sigma_{\rm bulk}\simeq130~\kms$. Even in this case, however, the emission component was faint, with a 3-$\sigma$ upper limit of $3.8\times10^{-5}$ photons s$^{-1}$ cm$^{-2}$, and improved the fit by only $\Delta C=6$ for $\Delta{\rm d.o.f.}=3$. We therefore conclude that a simple Fe~\textsc{xxvi} Ly$\alpha$ emission component is not statistically required and cannot explain the observed profile within this empirical test. A fully self-consistent photoionization treatment of absorption, re-emission, and scattered continuum components is beyond the scope of this Letter and will be addressed in future work. }

\begin{deluxetable*}{lccc}[t!] 
\tablecaption{Best-fit parameters for the phase-averaged and phase-resolved spectra.}
\tablehead{
\colhead{Parameter} & \colhead{Phase-averaged} & \colhead{Period $\phi_1$}  & \colhead{Period $\phi_2$}
}
\startdata
\sidehead{\bf Continuum Emission}$N_{\rm H}$ ($10^{22}\,\mathrm{cm^{-2}}$) & $24.3\pm1.0$ & \multicolumn{2}{c}{fixed to phase-averaged value} \\
$kT_{\rm bbody}$ (keV)                   & 2.00 (fixed) & \multicolumn{2}{c}{fixed to phase-averaged value} \\
Norm$_{\rm bbody}$                       & $6.2\pm0.1\times10^{-3}$ & \multicolumn{2}{c}{fixed to phase-averaged value}\\
$kT_{\rm disk}$ (keV)                    & 1.14 (fixed) & \multicolumn{2}{c}{fixed to phase-averaged value}\\
Norm$_{\rm disk}$                        & $24\pm1$ & \multicolumn{2}{c}{fixed to phase-averaged value}\\
Interstellar-absorption-corrected bolometric $L_X$ $(\mathrm{erg~s^{-1}})$ & $1.4\times10^{37}$ && \\
\sidehead{\bf Blueshifted Component}
$v_z$ (km s$^{-1}$)               & $-160\pm20$   & $-260\pm40$ & $-140^{+70}_{-80}$   \\
$\sigma _{\rm bulk}$ (km s$^{-1}$)       & $280\pm30$  & $150^{+40}_{-50}$ & $320^{+70}_{-50}$\\
$\sigma_{\rm turb}$ (km s$^{-1}$)                 & $100\pm10$ & \multicolumn{2}{c}{fixed to phase-averaged value}\\
Absorbed flux$_{2-10~\rm keV} (\mathrm{erg~cm^{-2}~s^{-1}})$ & $2.45^{+0.02}_{-0.03}\times10^{-10}$ && \\
Fe \textsc{xxvi} ($10^{18}$~cm$^{-2}$) & $5_{-2}^{+3}$ & $2.5^{+2.5}_{-1.2}$ & $12^{+10}_{-6}$  \\
Fe~\textsc{xxv} ($10^{18}$~cm$^{-2}$) & $1.4^{+0.4}_{-0.3}$  &&\\
Fe~\textsc{xxiv} ($10^{18}$~cm$^{-2}$)& $<0.06$~(3$\sigma$) &&\\
Ni~\textsc{xxviii} ($10^{18}$~cm$^{-2}$)& $0.7^{+0.3}_{-0.2}$ &&\\
Ni~\textsc{xxvii} ($10^{18}$~cm$^{-2}$) & $0.15^{+0.09}_{-0.06}$  &&\\
\sidehead{\bf Redshifted Component}
$v_z$ (km s$^{-1}$)           &  $+590^{+100}_{-80}$  & $+470\pm60$ & $+800^{+120}_{-110}$ \\
$\sigma _{\rm bulk}$ (km s$^{-1}$)   & \multicolumn{3}{c}{linked to blueshifted component value}    \\
$\sigma_{\rm turb}$ (km s$^{-1}$)  & \multicolumn{3}{c}{linked to blueshifted component value}    \\
Absorbed flux$_{2-10~\rm keV} (\mathrm{erg~cm^{-2}~s^{-1}})$ & $2.45^{+0.02}_{-0.03}\times10^{-10}$ && \\
Fe~\textsc{xxvi} ($10^{18}$~cm$^{-2}$)& $ 0.5^{+0.2}_{-0.1} $ & $0.6^{+0.3}_{-0.2}$ & $1.2^{+1.1}_{-0.5}$\\
Fe~\textsc{xxv} ($10^{18}$~cm$^{-2}$) & $<0.04$~(3$\sigma$)  &&\\
\hline
\sidehead{\bf Cross-contamination Factors} 
AX J1745 $\to$ MAXI J1744 region & $0.78\pm0.03$ &\\
MAXI J1744 $\to$ AX J1745 region & $0.817\pm0.007$ & \multicolumn{2}{c}{fixed to phase-averaged value}\\
\hline
C-statistic / d.o.f & 35782.96 / 33182 & \multicolumn{2}{c}{476.59 / 430}\\
\hline
\enddata
\tablecomments{
The Doppler velocity $v$ is tied within each component, while $\sigma$ in the Voigt profile is tied across all components. Errors are quoted at the 1$\sigma$ confidence level, and upper limits at a $3\sigma$ confidence level.
The \texttt{bbody} normalization is defined as $L_{39}/D_{10}^2$, where $L_{39}$ is the source luminosity in units of $10^{39}\,\mathrm{erg\,s^{-1}}$ and $D_{10}$ is the distance to the source in units of 10~kpc.
The \texttt{diskbb} normalization is $(R_{\rm in}/D_{10})^2 \cos\theta$, where $R_{\rm in}$ is the apparent inner disk radius in km and $\theta$ is the disk inclination angle ($\theta=0$ corresponds to a face-on disk).
}
\label{tab:bestfit}
\end{deluxetable*}

\textcolor{black}{Another possibility is an additional absorption component associated with \ax{}. We therefore introduced a second velocity component for Fe~\textsc{xxvi}, together with an Fe~\textsc{xxv} component tied to the same kinematic parameters. Because the photon statistics do not allow all line-profile parameters to be constrained independently for this weak redshifted component, we tied $\sigma_{\rm turb}$ and $\sigma_{\rm bulk}$ to those of the primary velocity component. As a result, the additional Fe~\textsc{xxvi} component, with $v=+590^{+100}_{-80}~\mathrm{km~s^{-1}}$ and $N_{\rm ion}=5^{+2}_{-1}\times10^{17}~{\rm cm^{-2}}$, explains the absorption-line profile shown in Figure~\ref{fig:spec_zoom} and significantly improves the fit statistic by $\Delta C=22$ for $\Delta {\rm d.o.f.}=2$. Only an upper limit was obtained for the Fe~\textsc{xxv} column density.}
\textcolor{black}{This component accounts not only for the inversion of the Fe~\textsc{xxvi} Ly$\alpha_1/\alpha_2$ depth ratio but also for the additional absorption trough around 6.93~keV. Such a separate trough was not reported for 4U~1624$-$49, where the anomaly was discussed primarily in terms of the doublet ratio \citep{trigo_2026}.}

We further tested the presence of intermediate velocity components or an asymmetric profile with velocities between the two components with $v=-160$ and $+590$~\kms. We introduced a third velocity component over the range from $v=-100$ to $+550~\mathrm{km~s^{-1}}$ in steps of $\Delta v=50~\mathrm{km~s^{-1}}$. For each trial, the velocity width was fixed to that of the other components, while only the column density was treated as a free parameter, and we examined whether the fit was statistically improved. As the result shown in the bottom panel in  Figure~\ref{fig:spec_zoom}, we confirmed that an intermediate velocity component did not improve the fit statistics. Therefore, we concluded that the best-fit model requires adding another velocity component of the Fe~\textsc{xxvi} ion to Equation \ref{eq:model}, which is expressed as follows:
\begin{equation}
\texttt{TBabs} * (\texttt{gsmooth} * A_1(v_1) * A_2(v_2)) * (\texttt{diskbb}+\texttt{bbody}),
\end{equation}
where
\begin{equation}
A_i(v_i) = \texttt{vashift}(v_i) * \prod \texttt{voigtabs}.
\end{equation}
Furthermore, the velocity separation between the two components ($\sim750~{\rm km~s^{-1}}$) significantly exceeds the turbulent and bulk broadening velocities ($\sigma_{\rm turb}=100\pm10$ and $\sigma_{\rm bulk}= 280 \pm30~{\rm km~s^{-1}}$), indicating that the two components are spectrally resolved and kinematically distinct. The best-fit parameters are listed in Table~\ref{tab:bestfit}. To assess the significance of the redshifted absorption component, when combining errors on the first component and photon noise, we computed Monte-Carlo (MC) simulations, following the method detailed in Appendix~\ref{app:MC}. Our simulations confirm a MC significance above p=0.999 ($\>3\sigma$).


\subsection{\textcolor{black}{Time Variability}}

\textcolor{black}{To investigate whether the kinematic structure of the absorption lines depends on orbital phase, we performed phase-resolved spectroscopy of the \ax{} region. We extracted the 5--12~keV light curve, where \ax{} dominates over the contaminating components (Figure~\ref{fig:spec}). Since the observation covers only about four orbital cycles, we adopted the 2024 XRISM orbital period, $P_{\rm orb}=30063$~s \citep{tanaka_2026}. The folded light curve shows a clear eclipse (Figure~\ref{fig:lc}); we define its center as $\phi_{\rm orb}=0$, exclude $\phi_0=0.96$--0.04, and divide the remaining out-of-eclipse data into two intervals, $\phi_1=0.04$--0.50 and $\phi_2=0.50$--0.96.}

\textcolor{black}{Because of the reduced photon statistics, we restricted the phase-resolved fits to the Fe~\textsc{xxvi} Ly$\alpha$ band, 6.91--7.01~keV. The Fe~\textsc{xxvi} column densities, velocities of the two absorption components, and $\sigma_{\rm bulk}$ were allowed to vary between $\phi_1$ and $\phi_2$, while $\sigma_{\rm turb}$ was fixed at the phase-averaged value because it is strongly degenerate with $\sigma_{\rm bulk}$ in this narrow-band fit and also affects the line optical depth; see Appendix~\ref{app:voi}. As a result, both phase intervals are consistent with a two-component velocity model, although the ionic column densities have large uncertainties and the fitted velocities and broadening suggest possible phase dependence (Table~\ref{tab:bestfit}).
}

\section{Discussion and Conclusion} \label{sec:dis}

\textcolor{black}{The principal result of this work is the resolution of two kinematically distinct components, one redshifted and one blueshifted, in the Fe~\textsc{xxvi} Ly$\alpha$ absorption profile in \ax{}. Since the continuum emission can be explained as a combination of blackbody and disk-blackbody emission, and exhibits prominent highly ionized absorption lines of Fe and Ni, \ax{} is likely to be in its soft state. Therefore, the absorption lines we detected can be interpreted in the context of a disk atmosphere or wind. We note that the systemic radial velocity of \ax{} is unconstrained. However, it would shift both components by the same amount and therefore does not affect their velocity separation or kinematic distinction.}

\textcolor{black}{We first consider the blueshifted component as a possible disk atmosphere or outflow. Adopting an orbital period of $P_{\rm orb}=$30063 s, a neutron-star mass of $M_{\rm NS}=1.4 M_\odot$, and a companion mass of \textcolor{black}{$M_{\rm comp}=0.4M_\odot$}, Kepler’s law gives a binary separation of \textcolor{black}{$a\simeq1.8\times10^{11}$} cm. The Roche-lobe radius of the neutron star is then \textcolor{black}{$R_{\rm L,NS}\simeq8.6\times10^{10}$} cm. Assuming that the disk extends to 0.7–-0.9$R_{\rm L,NS}$, as expected for a tidally truncated disk, we obtain the outer radius of the disk as \textcolor{black}{$R_{\rm out}\simeq6.0\text{--}7.7\times10^{10}$} cm. The corresponding escape velocity is \textcolor{black}{$v_{\rm esc}(R_{\rm out})\simeq690\text{--}790$} \kms. The velocity of the blueshifted component is therefore substantially below the escape velocity at the outer disk and is consistent with a slow outflow or disk-atmosphere material.
Such a slow blueshift, of order 100~\kms, is consistent with a thermal disk wind or disk-atmosphere component in a sub-Eddington soft state \citep{higginbottom_2019,higginbottom_2020}. Within this interpretation, the small observed blueshift may also imply that the systemic line-of-sight velocity of the system is also small.
Since this blueshifted component shows about ten times higher column densities than the redshifted one (Table~\ref{tab:bestfit}), the blueshifted component should have dominated the absorption profiles in the previous observations of \ax{}. The dominance of Fe~\textsc{xxvi} over Fe~\textsc{xxv} in the blueshifted component is consistent with the result from Suzaku \citep{hyodo_2009} and \xmm{} \citep{ponti_2015}, suggesting that this system persistently hosts a highly ionized disk atmosphere or outflow during the soft state.}

\textcolor{black}{The redshifted component is difficult to explain as either a nearly static disk atmosphere or an outflow. Its phase-averaged velocity, $v=+590^{+100}_{-80}$~\kms, is somewhat lower than, but of the same order as, the escape velocity at the outer disk radius, $v_{\rm esc}=690\text{--}790$~\kms. This is compatible with a failed-wind scenario, in which material launched from the disk fails to escape and subsequently falls back toward the neutron star. If so, the present observation may provide a snapshot of both outflowing and returning material in the disk-wind system. A redshifted component with a similar velocity was reported in the black-hole LMXB MAXI~J1305$-$704 \citep[$+540$~\kms;][]{miller_2014} and was interpreted as fallback from a failed wind. Infalling or failed-wind signatures have also been reported at optical and infrared wavelengths in GRS~1716$-$249 \citep{cuneo_2020}, where inverse P-Cygni profiles of H$\alpha$, H$\beta$, and He~\textsc{ii} showed redshifted absorption reaching approximately 1300~\kms.}

\textcolor{black}{Another possible origin is a gravitational redshift. When the absorber is located at $R_{\rm abs}$ from the NS, the gravitational redshift is $z=GM_{\rm NS}/c^2R_{\rm abs}$. In addition, the transverse Doppler effect contributes an additional $z_{\rm TD} = GM_{\rm NS}/2c^2R_{\rm abs}$, giving a total redshift of $z_{\rm total} = 3GM_{\rm NS}/2c^2R_{\rm abs}$ \citep{trueba_2022}. Therefore, if the redshift velocity in \ax\ originates from a gravitational redshift of ions moving in the Keplerian orbit, the radial position of the absorber is
\begin{eqnarray}
    R_{\rm abs}
    &=&7.6\times10^2R_g\left( \frac{v_{\rm shift}}{590~{\rm km~s^{-1}}}\right)^{-1}\nonumber\\
    &=&1.6\times 10^8~{\rm cm}~\left( \frac{M_{\rm NS}}{1.4~M_\odot}\right)\left( \frac{v_{\rm shift}}{590~{\rm km~s^{-1}}}\right)^{-1},
    \label{eq:grr}
\end{eqnarray}
where $R_g=GM_{\rm NS}/c^2$ is the gravitational radius.
For comparison, the disk-blackbody normalization gives a color-corrected inner disk radius of approximately $R_{\rm in}=2\times10^6$ cm for $d=8$ kpc, $i=70^\circ$, and $f_{\rm col}=1.7$. 
This estimate depends on the fixed disk temperature approximately as $R_{\rm in}\propto kT_{\rm disk}^{-2}$, but this uncertainty does not affect the conclusion that $R_{\rm abs}$ is larger than $R_{\rm in}$.
\citet{trueba_2022} also reported a signature of the redshifts of highly-ionized Fe ions in \ax{} with the velocity ($v=270^{+240}_{-230}$~\kms) using \chandra{}/HETG, and they discussed the possible origins from gravitational redshifts. Although this previous measurement does not overlap with our phase-averaged velocity, $v=+590^{+100}_{-80}~{\rm km~s^{-1}}$, the difference is not surprising given the decade-long separation between the observations.}

\textcolor{black}{The redshifted component is also seen in both phase intervals, $\phi_1$ and $\phi_2$. This behavior stands in contrast to the redshifted absorption components in other dippers, such as 4U~1630$-$472 \citep{miller_2025}, where the feature becomes deeper specifically during dipping phases. The presence of the redshifted component in both broad out-of-eclipse intervals disfavors an origin confined exclusively to a localized structure such as the stream-impact bulge. At the same time, the absorption profile in $\phi_2$ is broader than in $\phi_1$. Such phase-dependent broadening is difficult to attribute to orbital motion alone, which would shift the line centroid rather than increase the velocity width, and instead suggests that the dominant absorbing material may be azimuthally non-uniform, as inferred for other eclipsing systems \citep[e.g., EXO 0748$-$676; ][]{parmar_1986}. }

\textcolor{black}{
Both an infalling flow and gravitational redshift remain physically plausible, and neither can be ruled out with the present data.
An inverse P-Cygni profile would provide a decisive signature of inflowing material; such profiles have been detected at optical and infrared wavelengths in GRS~1716$-$249 and interpreted as evidence for infall \citep{cuneo_2020}. Although the limited photon statistics of the present observation do not allow us to meaningfully constrain such a profile, future on-axis XRISM observations combined with self-consistent photoionization modeling may provide a sensitive test of this scenario. Regardless of whether the redshifted component arises from failed-wind fallback or gravitational redshift, the Fe~\textsc{xxvi} Ly$\alpha$ profile requires two well-separated concentrations of absorbing optical depth in velocity space, with no significant absorption detected at intermediate velocities. This disfavors a single homogeneous absorber with a smooth, monotonic velocity field and instead requires at least two kinematically distinct absorbing zones, or a strongly non-monotonic flow. In the failed-wind interpretation, the result suggests that outflowing and returning material occupy distinct flow channels or episodic structures rather than forming a smooth velocity continuum. In the gravitational-redshift interpretation, it requires an inner absorbing zone spatially separated from the dominant outer disk atmosphere or wind. Thus, the discrete velocity structure revealed by Resolve provides direct evidence that the highly ionized disk atmosphere in \ax{} is dynamically stratified.}

\section{Acknowledgement}
\nolinenumbers
\textcolor{black}{We thank the anonymous referee for insightful and constructive comments that substantially improved the clarity and physical interpretation of the data, particularly by motivating the phase-resolved analysis and a broader discussion.}
We also thank the \xrism{} operation team for accepting our DDT proposal and conducting the observation, along with the \xrism{} Science Data Center, help desk, and calibration teams for their continued assistance. We thank Koh Sakamoto, Hiroyuki Uchida, Takeshi Go Tsuru, Yuken Ohshiro, Shinya Yamada, and Shogo B. Kobayashi for their valuable comments on our analysis and discussion. This work was supported by the JSPS KAKENHI grants number 24KJ1485, 19K14762, 23K03459, 24H01812, J24KF0244, 24KJ0152, 26H02075, and 23H01211. MP acknowledges support from the JSPS Postdoctoral Fellowship for Research in Japan, grant number P24712.  Support for SM, KM and the Columbia University team was provided by \nustar\ AO-10 (80NSSC26K0286), \nustar\ AO-11 (80NSSC26K0154), \chandra\ AO-26 (SAO GO5-26016X) and \xmm\ AO-23  (80NSSC25K0651) programs. 
SM acknowledges support by the National Science Foundation Graduate Research Fellowship under Grant No. DGE 2036197 and the Columbia University Provost Fellows Program.  

\appendix

\section{Response Generation} \label{app:arf}

\textcolor{black}{XRISM Auxiliary Response Files (ARFs) are generated using \texttt{xaxmaarfgen}, which utilizes the \texttt{xrtraytrace} ray-tracing simulator along with an exposure map and the XMA Calibration Database (CalDB). The exposure map is pre-generated by \texttt{xaexpmap} using the CalDB and the spacecraft attitude data, and it records the optical axis offset keywords (\texttt{OPTAOFFX} and \texttt{OPTAOFFY}). These tools (\texttt{xaxmaarfgen}, \texttt{xrtraytrace}, and \texttt{xaexpmap}) are provided as part of HEASoft. In HEASoft v6.36, \texttt{xaxmaarfgen} is known to use an incorrect region in the simulated image when calculating the effective area if run with CalDB v12. To circumvent this issue, we set both optical axis offset keywords to 0.0, following the procedure recommended by the XRISM Science Data Center \footnote{https://heasarc.gsfc.nasa.gov/docs/xrism/analysis/ttwof/index.html}.}

\section{Background Emission Model} \label{app:bgd}

In this study, we also modeled the emission from \maxi, Sgr A East, and GCXE. For \maxi, we followed the model of \citet{parra_2026}, studying the same \xrism{} observation together with simultaneous observations, combining with \nustar{} and \xmm{}. The model is
\begin{equation}
\texttt{TBabs}(\texttt{ThComp}(\texttt{diskbb})+\texttt{vashift}(\Sigma \texttt{Gaussian})),
\end{equation}
where \texttt{TBabs} is an interstellar absorption model \citep{wilms_2000}, \texttt{diskbb} is blackbody emission from the accretion disk, \texttt{ThComp} is comptonized photon components by thermal electrons. They modeled the emission lines from BHLMXB by Gaussians, except for Fe I K$\alpha$ and Fe I K$\beta$, modeled by Holzer profiles \citep{holzer_1997}. See \citet{parra_2026} for details and parameters.

Sgr A East is a supernova remnant located near the FoV. \citet{xrismcollaboration_2025} performed the spectroscopy with the \xrism{}/Resolve, and we used their model as the background for our analysis:
\begin{equation}
\texttt{TBabs}(\texttt{bvvrnei}+\texttt{bvvrnei}+\texttt{power-law}+(\Sigma~\texttt{Lorentz})).
\end{equation}
\texttt{TBabs} is the interstellar absorption with $N_{\rm H}=1.5\times10^{23}~{\rm cm^{-2}}$. Two \texttt{bvvrnei} are over-ionized plasma components, which mainly represent the emission from ionized Fe.
The continuum emission is represented with a \texttt{power-law} component. The Lorentzian components correspond to emission lines from Fe~\textsc{I}. The redshift and Gaussian-like broadening of the lines emission components were reported as $z=(2\pm5)\times10^{-5}$ (i.e., $6\pm15$~km~s$^{-1}$) and $\sigma=109\pm6~{\rm~km~s^{-1}}$, respectively, while the fluxes of Fe \textsc{xxvi} K$\alpha_1$ and K$\alpha_2$ at the source region were $4.9\pm0.8\times10^{-6}$ and $3.3\pm0.7\times10^{-6}$ ph~s$^{-1}$~cm$^{-2}$, respectively. 

The Galactic center X-ray emission (GCXE) is a thermal emission distributed along the Galactic plane \cite[][for review]{koyama_2018}, which can be modeled with hot plasma components \citep{uchiyama_2013}. In order to estimate the contribution to the \ax{} region, we refer to the model used in the previous study on Sgr A East supernova remnant \citep{xrismcollaboration_2025}, where the authors empirically modeled the spectrum from a nearby-sky observation with a power-law continuum component and two thermal plasma components with neutral iron and nickel lines. We also only left the surface brightness as the free parameter, then we fitted the 2024 observations of the region around \ax{} jointly with the Sgr A East and \ax{} components, and estimated the surface brightness. 
Then we incorporated the GCXE component by fixing its parameters and brightness to the values obtained from the 2024 data. As shown in Figure 2, the contribution of the GCXE is smaller than that of Sgr A East, and its uncertainty is thus unlikely to affect our conclusions.

\section{Voigtabs Modeling} \label{app:voi}

\begin{table}
\centering
\footnotesize
\caption{Atomic data of absorption lines included in the spectral model}
\label{tab:atomic}
\begin{tabular}{lccc}
\hline
Ion & Line ID & $E_{\rm c}$ (keV) & $f_{ik}$ \\
\hline
Fe~\textsc{xxvi} & Ly$\gamma_1$ & 8.701 & $1.9\times10^{-2}$ \\
Fe~\textsc{xxvi} & Ly$\gamma_2$ & 8.699 & $9.4\times10^{-3}$ \\
Fe~\textsc{xxvi} & Ly$\beta_1$  & 8.253 & $5.2\times10^{-2}$ \\
Fe~\textsc{xxvi} & Ly$\beta_2$  & 8.246 & $2.6\times10^{-2}$ \\
Fe~\textsc{xxvi} & Ly$\alpha_1$ & 6.973 & $2.7\times10^{-1}$ \\
Fe~\textsc{xxvi} & Ly$\alpha_2$ & 6.952 & $1.4\times10^{-1}$ \\
\\
Fe~\textsc{xxv}  & He$\gamma_1$  & 8.295 & $5.1\times10^{-2}$ \\
Fe~\textsc{xxv}  & He$\gamma_2$  & 8.292 & $6.0\times10^{-3}$ \\
Fe~\textsc{xxv}  & He$\beta_1$   & 7.881 & $1.4\times10^{-1}$ \\
Fe~\textsc{xxv}  & He$\beta_2$   & 7.872 & $1.7\times10^{-2}$ \\
Fe~\textsc{xxv}  & He$\alpha$ (w)& 6.700 & $7.0\times10^{-1}$ \\
Fe~\textsc{xxv}  & He$\alpha$ (x)& 6.682 & $1.7\times10^{-5}$ \\
Fe~\textsc{xxv}  & He$\alpha$ (y)& 6.668 & $6.9\times10^{-2}$ \\
Fe~\textsc{xxv}  & He$\alpha$ (z)& 6.637 & $3.3\times10^{-7}$ \\
\\
Fe~\textsc{xxiv} &  & 6.679 & $5.0\times10^{-1}$ \\
Fe~\textsc{xxiv} &  & 6.676 & $1.6\times10^{-1}$ \\
Fe~\textsc{xxiv} &  & 6.662 & $4.6\times10^{-3}$ \\
Fe~\textsc{xxiv} &  & 6.653 & $9.9\times10^{-2}$ \\
Fe~\textsc{xxiv} &  & 6.617 & $1.6\times10^{-2}$ \\
Fe~\textsc{xxiv} &  & 6.613 & $2.2\times10^{-3}$ \\
\\
Ni~\textsc{xxviii} & Ly$\beta_1$  & 9.586 & $5.1\times10^{-2}$ \\
Ni~\textsc{xxviii} & Ly$\beta_2$  & 9.578 & $2.6\times10^{-2}$ \\
Ni~\textsc{xxviii} & Ly$\alpha_1$ & 8.102 & $2.7\times10^{-1}$ \\
Ni~\textsc{xxviii} & Ly$\alpha_2$ & 8.073 & $1.4\times10^{-1}$ \\
\\
Ni~\textsc{xxvii} & He$\gamma_1$  & 9.667 & $4.7\times10^{-2}$ \\
Ni~\textsc{xxvii} & He$\gamma_2$  & 9.662 & $7.4\times10^{-3}$ \\
Ni~\textsc{xxvii} & He$\beta_1$   & 9.184 & $1.3\times10^{-1}$ \\
Ni~\textsc{xxvii} & He$\beta_2$   & 9.172 & $2.0\times10^{-2}$ \\
Ni~\textsc{xxvii} & He$\alpha$ (w)& 7.806 & $6.8\times10^{-1}$ \\
Ni~\textsc{xxvii} & He$\alpha$ (x)& 7.786 & $2.3\times10^{-6}$ \\
Ni~\textsc{xxvii} & He$\alpha$ (y)& 7.766 & $8.8\times10^{-2}$ \\
Ni~\textsc{xxvii} & He$\alpha$ (z)& 7.732 & $5.2\times10^{-7}$ \\
\hline
\end{tabular}
\tablecomments{
$E_{\rm c}$ denotes the rest-frame transition energy and $f_{ik}$ is the oscillator strength.
Atomic data are taken from the NIST Atomic Spectra Database \citep{nist_asd_2024}.
}
\end{table}

We modeled the line absorptions in \ax\ with \texttt{voigtabs} in Xspec, which assumes an absorber whose absorption cross section as a function of energy follows a Voigt profile:
\begin{equation}
F(E) = F_0(E)\,\exp[-\tau(E)],
\label{eq:vabs}
\end{equation}
with
\begin{equation}
    \tau(E) = d\,v(E-E_c;\Gamma,\sigma),
\end{equation}
where $d$ is the line depth in keV unit and $v(E-E_c, \Gamma, \sigma)$ is a normalized Voigt function normalized to 1 in eV dimension. In case of line absorptions of ions, we can assume that the $\Gamma$ and $\sigma$ correspond to natural widths of the transitions and turbulent broadening velocity. The constant $d$ satisfies the following equation:
\begin{equation}
    d=N_{\rm ion}\frac{\pi e^2}{m_{\rm e}c}f_{ik}
    \label{eq:tau}
\end{equation}
Here, $N_{\rm{ion}}$ is the column density of the ion, $e$ and $m_{\mathrm e}$ are the electron charge and mass, $c$ is the speed of light, and $f_{ik}$ is the oscillator strength of the transition. In this study, $\Gamma$ and $f_{ik}$ of all transitions are taken from the  National Institute of Standards and Technologies Atomic Spectra Database (NIST ASD) Version 5.12 \citep{nist_asd_2024}. When assuming a series of absorption lines produced by a single ion species, one of the line-strength parameters $d$ can be treated as a free parameter, while all the remaining $d$ values are tied to it through the ratios of the oscillator strengths $f_{ik}$. The ions we modeled in this study and their oscillator strengths are listed in Table~\ref{tab:atomic}. Assuming the $\sigma_{\rm turb}$ should be shared among different transitions in a single ion species, the series of absorption lines can then be described (at first order) using two free parameters: $d$ and $\sigma_{\rm turb}$. Once we obtained a $d$ parameter of any transition, we can directly get the $N_{\rm ion}$ based on Equation~\ref{eq:tau}.

\section{Monte-Carlo simulations}\label{app:MC}
To consider the look elsewhere effect and derive the 
significance of the redshifted component, we computed Monte-Carlo simulations, as detailed in \cite{parra_2026a} for MAXI J1744-294. Namely, from the base fit of our "null-hypothesis" model without the redshifted component, we simulated 1000 Resolve spectra using the \texttt{fakeit} command in xspec, allowing statistical simulations. Then, for each simulated spectrum, after refitting with the null-hypothesis model to derive a baseline C-statistic, we scanned the parameter space of our redshifted component for improvements to the fit due to photon noise. The maximum C-statistic improvement in the allotted parameter space, $\Delta C_{\rm fake}$, can then be compared to the statistical significance of the ``real'' feature, $\Delta C_{\rm real}$, in each of the 1000 simulated spectra. This yield a significance $p=1-N/1000$, with N the number of simulated spectra where $\Delta C_{\rm real}<\Delta C_{\rm fake}$.

Since the velocity shift of the redshifted component is small enough to unambiguously attribute it to Fe~\textsc{xxvi} Ly$\alpha$, we followed the approach of \cite{parra_2026a} and scanned the $[-3000,3000]$ km s$^{-1}$ velocity band around the rest energy of the Fe~\textsc{xxvi} Ly$\alpha$ transitions. However, the \ax{} configuration is more complex than that of \maxi, due to the presence of the main absorption component and its two transitions, which not only influence the fitting of the redshifted component due to their overlap, but significantly lowers count rate in its energy range, which could increase the impact of photon noise. To consider simultaneously the effect of the photon noise and the fitting uncertainty on the main component, we thus derived 1000 simulated set of parameters of the main absorption component from its posterior distribution, computed from the one-component best fit using a Monte-Carlo chain. These simulated line parameters were then used as the base models for each of the \texttt{fakeit} runs. In each simulation,  the baseline model (notably including the lines) was refitted, before adding another Fe~\textsc{xxvi} Ly$\alpha$ duet with a $2-1$ normalization radio. Finally, for searching improvements in C-statistic, a \texttt{steppar} exploration of the velocity shift is generally considered sufficient, but not here, due to the number of free parameters and the inherent degeneracy between the two complexes. Instead, we thus manually fitted the two components together while varying the velocity of the second component in increments of 25 km s$^{-1}$, spanning the whole velocity band defined previously. It is worth noting that the photon noise improvements are largely dominated by the $[-1000,1000]$ km s$^{-1}$ band where the spectrum is significantly absorbed by the main Fe~\textsc{xxvi} Ly$\alpha$ component, and the results of this test would thus remain identical even for a velocity band several times larger.



\bibliography{2026_AXJ1745, other}{}
\bibliographystyle{aasjournal}

\end{document}